\documentclass[aps,graphicx,twocolumn]{revtex4}

\usepackage{graphicx}
\usepackage{epstopdf}
\usepackage{verbatim}
\usepackage{amsmath}
\usepackage[colorlinks,linkcolor=blue,citecolor=blue,urlcolor=blue]{hyperref}
\begin{document}
\title{Memory-assisted measurement-device-independent quantum secret sharing}

\author{Cheng Zhang$^{1}$, Qi Zhang$^{2}$, Wei Zhong$^{3}$, Ming-Ming Du$^{1}$, Shu-Ting Shen$^{1}$, Xi-Yun Li$^{2}$, An-Lei Zhang$^{2}$, Lan Zhou$^{2}$\footnote{Email address: zhoul@njupt.edu.cn}, Yu-Bo Sheng$^{1,3}$}
\address{
$^1$College of Electronic and Optical Engineering, \& College of Flexible Electronics (Future Technology), Nanjing University of Posts and Telecommunications, Nanjing, 210023, China\\
$^2$College of Science, Nanjing University of Posts and Telecommunications, Nanjing,\\
$^3$Institute of Quantum Information and Technology, Nanjing University of Posts and Telecommunications, Nanjing, 210003, China\\
}
\date{\today}

\begin{abstract}
Measurement-device-independent quantum secret sharing (MDI-QSS) can eliminate all the security loopholes associated with imperfect measurement devices and greatly enhance QSS's security under practical experimental condition. MDI-QSS requires each communication user to send single photon to the measurement party for the coincident measurement. However, the unsynchronization of the transmitted photons greatly limits MDI-QSS's practical performance. In the paper, we propose a high-efficient quantum memory (QM)-assisted MDI-QSS protocol, which employs the QM-assisted synchronization of three heralded single-photon sources to efficiently generate three simultaneous single-photon states. The QM constructed with all-optical, polarization-insensitive storage loop has superior performance in terms of bandwidth, storage efficiency, and noise resistance, and is feasible under current experimental conditions.
Combining with the decoy-state method, we perform the numerical simulation of the secure key rate in the symmetric model without considering the finite-size effect. The simulation results show that
our protocol has largely improved secure key rate and maximal photon transmission distance compared with all existing MDI-QSS protocols without QM.
Our protocol provides a promising way for implementing the high-efficient MDI-QSS in the near future.\\
\end{abstract}
\maketitle

\section{Introduction}
Quantum communication has the unconditional security in principle based on the basic principle of quantum mechanics, and attracts much attention. It starts from the research on quantum key distribution (QKD) in 1984, which can generate random keys between two distant communication users \cite{QKD1,QKD2}. In the past 40 years, QKD has achieved great development and has become the most practical quantum communication technology \cite{QKD3,QKD4,QKD5,QKD8,QKD9}. Besides QKD, quantum communication also includes some important branches, such as quantum secure direct communication (QSDC) \cite{QSDC1,QSDC2,QSDC6} and quantum secret sharing (QSS) \cite{QSS1,QSS2,QSS3}.
QSS is an important multipartite cryptographic primitive. It splits each key of the dealer into several pieces and distributes each piece to a player. Any subset of players cannot reconstruct the distributed keys, which can be reconstructed only when all the players cooperate \cite{QSS1,QSS2,QSS3}. Conversely, all the players can also cooperate to distribute a secure key to the dealer.

QSS was originally proposed by Hillery \emph{et al.} in 1999 \cite{QSS1}, which bases on the quantum technology and the traditional cryptographic sharing technology. Since then, QSS has been widely researched in theory and experiment. The QSS protocols based on the Greenberger-Horne-Zeilinger (GHZ) state, Bell state, single photons, and coherent states have been successively proposed \cite{QSS5,QSS6,QSS7,QSS8,QSS9,QSS10,QSS11,QSS12,QSS13,QSS14,QSS15,QSS16}. Recently, researchers proposed the differential phase shift (DPS) QSS \cite{DPS2} and round-robin (RR) QSS \cite{RR1} protocols. In the experimental aspect, the proof-of-principle experimental demonstration of QSS based on entanglement \cite{QSSe2,QSSe4,QSSe6,QSSe6n}, single qubit \cite{QSSe3}, graph state \cite{QSSe5,QSSe5n} and coherent state \cite{QSSe7} have been reported.

 Similar to other quantum communication branches, although QSS has unconditional theoretical security, the practical imperfect experimental devices may cause security loopholes. For example,
  the eavesdropper (Eve) may perform the photon number splitting (PNS) attack \cite{PNS}, detection blinding attack \cite{blind1,blind2} and time-shift attack \cite{Time}. Measurement-device-independent (MDI) quantum communication protocols can resist all possible attacks from the imperfect measurement devices \cite{MDI1,MDI3,MDI4,MDI5,MDI12,MDIQSS1,MDIQSS2}. In 2015, the first MDI-QSS protocol combined with the decoy-state method \cite{MDI4} was proposed, which can guarantee the key security under practical experimental conditions.

  In the MDI-QSS protocol, the users require to send single photons to the measurement party for the GHZ state measurement. However, current available single-photon source is the phase-randomized weak coherent pulse
(WCP) source, which can emit vacuum state, single-photon state, and multi-photon state with different probabilities. Meanwhile, the channel noise may cause the photon transmission loss. The imperfect photon source and photon transmission loss lead to quite low coincidence counting rate in the GHZ state measurement, which largely limit MDI-QSS's secure key rate and photon transmission distance.
Similarly, the unsynchronization problem of the transmitted photons also exists in MDI-QKD protocols. In the MDI-QKD field, for solving the unsynchronization problem, some researchers proposed the QM-assisted MDI-QKD protocols to realize the two-photon synchronous projection measurement  \cite{QMMDI1,QMMDI2,QMMDI3}. Meanwhile, one possible approach to solve the imperfect photon source problem is to use the heralded single-photon  source (HSPS) \cite{HSPS1,HSPS3,HSPS4,HSPS5,HSPS6}. Suppose that a spontaneous parametric down-conversion (SPDC) source emits correlated photon pairs in two spatial modes. The detection of the photons in one of the two correlated spatial modes can herald the photon number statistics of the other spatial mode. This approach can significantly reduce the probability of the vacuum state and thus increase the coincidence counting rate of the GHZ state measurement. In 2017, Kaneda \emph{et al.}  proposed a QM-assisted HSPS-MDI-QKD protocol. Based on the herald property of HSPS, the photon arriving at the measurement side is firstly stored in the QM to wait for other photons to arrive, and will be released until the later photon arrives \cite{yuanshi}. Later, the QM-assisted HSPS-MDI-QKD combined with the decoy-state method was proposed in 2023, which can further resist the PNS attack \cite{wang}. Based on previous researches, the employment of HSPS and QM into MDI-QSS is a promising method to improve its performance under practical experimental condition.

In the paper, we propose a QM-assisted MDI-QSS protocol combining with the HSPS and the decoy-state method. Our protocol employs three QMs to synchronize three HSPSs to efficiently generate three simultaneous single photons. The QM constructed with the all-optical, polarization-insensitive storage loop displays superior bandwidth, storage efficiency, and noise resistance performances. Moreover, it is feasible under current experimental condition. Benefitting from the QMs, our protocol can efficiently increase the coincidence counting rate of the GHZ state measurement. Combining with the decoy-state method, our protocol can guarantee the security of the transmitted keys under practical experimental conditions. We develop numerical methods to simulate its secure key rate  in practical communication situation. Comparing with existing WCP-MDI-QSS and HSPS-MDI-QSS protocols without the QM, our QM-assisted MDI-QSS protocol has higher secure key rate and longer photon transmission distance. Based on above advantages, our protocol has potential to realize high-efficient MDI-QSS in the near future.

The paper is organized as follows. In Sec. \uppercase\expandafter{\romannumeral2}, we introduce our QM-assisted MDI-QSS protocol combined with the decoy-state method. In Sec. \uppercase\expandafter{\romannumeral3}, we establish a theoretical simulation model of the secure key rate of our QM-assisted MDI-QSS protocol. Finally, we make some discussion and draw a conclusion in Sec. IV.

\section{The QM-assisted MDI-QSS protocol}
\subsection{Key knowledge of the QM-assisted MDI-QSS protocol}
Before explaining our QM-assisted MDI-QSS protocol, we firstly introduce the following key knowledge. Each of the three communication users Alice, Bob and Charlie needs to use the rectilinear $(Z)$ basis and diagonal $(X)$ basis to generate single photons. In detail, we can describe $Z$ basis and $X$ basis as
\begin{eqnarray}
  &Z = \{|H\rangle,|V\rangle \},\nonumber\\
   &X=\{|+\rangle=\frac{1}{\sqrt{2}}(|H\rangle+|V\rangle), |-\rangle=\frac{1}{\sqrt{2}}(|H\rangle-|V\rangle)\},
\end{eqnarray}
where $|H\rangle$ and $|V\rangle$ represent the horizontal and vertical polarization states, respectively.

The eight GHZ states in $Z$ basis can be written as
\begin{eqnarray}
    &|\Phi^{\pm}\rangle=\frac{1}{\sqrt{2}}(|HHH\rangle\pm|VVV\rangle),\nonumber\\
    &|\Phi_1^{\pm}\rangle=\frac{1}{\sqrt{2}}(|VHH\rangle\pm|HVV\rangle),\nonumber\\
    &|\Phi_2^{\pm}\rangle=\frac{1}{\sqrt{2}}(|HVH\rangle\pm|VHV\rangle),\nonumber\\
    &|\Phi_3^{\pm}\rangle=\frac{1}{\sqrt{2}}(|HHV\rangle\pm|VVH\rangle).\label{GHZ}
\end{eqnarray}
\begin{figure}[htbp]
 \begin{center}
        \includegraphics[width=9cm,angle=0]{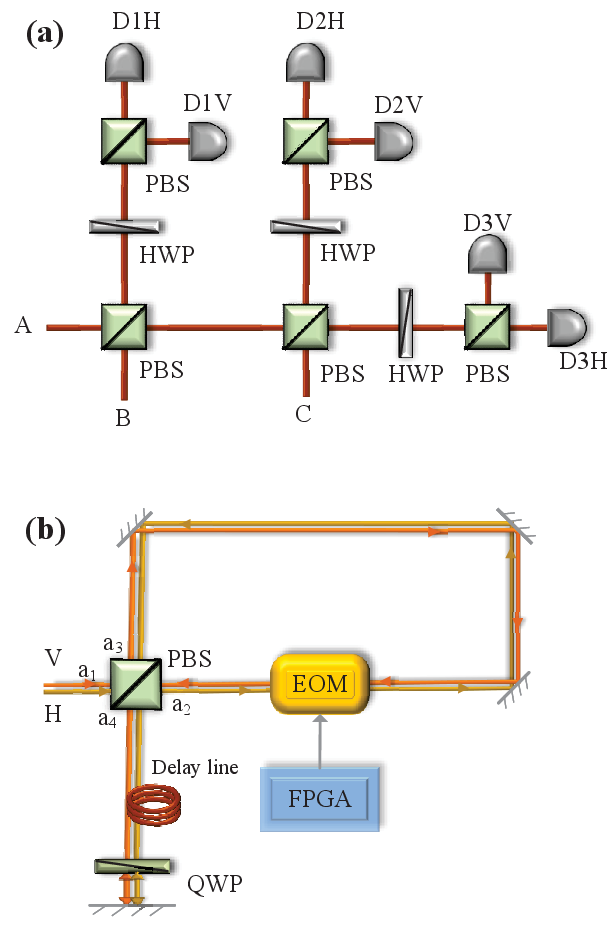}
        \caption{(a) The structure diagram of the GHZ state measurement module with linear optical elements. The measurement module can only distinguish two GHZ states $|\Phi^{\pm}\rangle$ in Eq. (\ref{GHZ}) \cite{GHZ}. (b) The structure diagram of the QM module. The QM can control the storage and readout of the photons by controlling on-off of the electro-optic modulator (EOM). When the EOM is turned on, the polarization of the passing photon will be rotated by 90 degrees.  The polarization beam splitter (PBS) can totally transmit the horizontally polarized photon and totally reflect the vertically polarized photon. HWP and QWP represent the half-wave plate and quarter wave plate, respectively. D1H, D1V, D2H, D2V, D3H and D3V are six single-photon detectors. The field programmable gate array (FPGA) is used as the signal control system for the QM. }
        \label{fig:1}
 \end{center}
\end{figure}

The GHZ state measurement module in our  protocol is composed of some linear optical elements as shown in the Fig. \ref{fig:1}(a) \cite{GHZ}.
The  measurement module can only distinguish two of the eight GHZ states $|\Phi^{\pm}\rangle=\frac{1}{\sqrt{2}}(|HHH\rangle\pm|VVV\rangle)$. The detector-click combination $D_{1H}D_{2H}D_{3H}$, $D_{1H}D_{2V}D_{3V}$, $D_{1V}D_{2H}D_{3V}$, or $D_{1V}D_{2V}D_{3H}$ corresponds to the state  $|\Phi^{+}\rangle$, while the detector-click combination $D_{1V}D_{2V}D_{3V}$, $D_{1H}D_{2H}D_{3V}$, $D_{1V}D_{2H}D_{3H}$, or $D_{1H}D_{2V}D_{3H}$ corresponds to the state $|\Phi^{-}\rangle$.
  $|\Phi^{\pm}\rangle$ can be transformed in $X$ basis as
  \begin{eqnarray}
  |\Phi^{+}\rangle&=&\frac{1}{2}(|+++\rangle+|+--\rangle+|-+-\rangle+|--+\rangle),\nonumber\\
  |\Phi^{-}\rangle&=&\frac{1}{2}(|---\rangle+|++-\rangle+|-++\rangle+|+-+\rangle).\nonumber\\
\end{eqnarray}

To facilitate understanding how users use the result of GHZ state measurement to generate keys, we show the correlation between the GHZ state measurement results and the input single photon states in Tab. \ref{tab:1}. From Tab. \ref{tab:1}, if all the three photons from Alice, Bob and Charlie are encoded in  $Z$ basis and the GHZ state measurement is successful, users can get $Z_A=Z_B=Z_C$. When all the three photons from Alice, Bob and Charlie are encoded in $X$ basis, if the measurement result is $|\Phi^+\rangle$, users can get $X_A=X_B\oplus X_C$ and if the measurement result is $|\Phi^-\rangle$, users can get $X_A\oplus1=X_B\oplus X_C$.
\begin{table}[htbp]
    \centering
    \vspace{-0.3cm}
    \setlength{\abovecaptionskip}{0.3cm}
    \setlength{\belowcaptionskip}{0.1cm}
    \setlength\tabcolsep{10pt} 
    \renewcommand\arraystretch{1.5}  
    \caption{The relationship between the GHZ state measurement results and the input single photon states.}
    \begin{tabular}{cccccc}
        \hline \hline
        \multicolumn{3}{c}{Input single photon state} &\multicolumn{2}{c}{ Probability of GHZ } \\
        \multicolumn{3}{c}{}  &\multicolumn{2}{c}{ state measurement result} \\
        \hline
        Alice&Bob&Charlie&$|\Phi^{+}\rangle$&$|\Phi^{-}\rangle$\\
       \hline
       $|H\rangle$ &$|H\rangle$ &$|H\rangle$ &1/2 &1/2 \\
       $|H\rangle$ &$|H\rangle$ &$|V\rangle$ &0 &0\\
       $|H\rangle$ &$|V\rangle$ &$|H\rangle$ &0 &0 \\
       $|V\rangle$ &$|H\rangle$ &$|H\rangle$ &0 &0 \\
       $|H\rangle$ &$|V\rangle$ &$|V\rangle$ &0 &0 \\
       $|V\rangle$ &$|H\rangle$ &$|V\rangle$ &0 &0 \\
       $|V\rangle$ &$|V\rangle$ &$|H\rangle$ &0 &0 \\
       $|V\rangle$ &$|V\rangle$ &$|V\rangle$ &1/2 &1/2 \\
       $|+\rangle$ &$|+\rangle$ &$|+\rangle$ &1 &0\\
       $|+\rangle$ &$|-\rangle$ &$|-\rangle$ &1 &0\\
       $|-\rangle$ &$|+\rangle$ &$|-\rangle$ &1 &0\\
       $|-\rangle$ &$|-\rangle$ &$|+\rangle$ &1 &0\\
       $|-\rangle$ &$|+\rangle$ &$|+\rangle$ &0 &1\\
       $|+\rangle$ &$|-\rangle$ &$|+\rangle$ &0 &1\\
       $|-\rangle$ &$|+\rangle$ &$|+\rangle$ &0 &1\\
       $|-\rangle$ &$|-\rangle$ &$|-\rangle$ &0 &1\\
        \hline
        \hline
        \label{tab:1}
    \end{tabular}
\end{table}

In our QM-assisted MDI-QSS protocl, three feasible all-optical, polarization-insensitive storage loops \cite{QM}  are employed as the QMs to assist the photon synchronization. The structure of QM is shown in Fig. \ref{fig:1}(b). Here, suppose a photon in $|H\rangle$ in $a_{1}$ mode enters the storage loop from the polarization beam splitter (PBS), which can totally transmit the photon in $|H\rangle$ to $a_{2}$ mode and totally reflect the photon in $|V\rangle$ to $a_{3}$ mode. EOM locating in $a_{2}$ mode is a bidirectional electro-optic modulator, which is controlled by the field programmable gate array (FPGA). When EOM is turned on, the polarization state of the single photon passing through EOM will be flipped ($|H\rangle\xrightarrow{EOM(ON)}|V\rangle$, $|V\rangle\xrightarrow{EOM(ON)}|H\rangle$). When EOM is turned off, the polarization of the passing photon will not change. The combination of the quarter wave plate (QWP) and the mirror in $a_{4}$ mode can rotate the polarization of the single photon coming out of the $a_4$ port by 90 degrees, that is, $|H\rangle\xrightarrow{double\, QWP}|V\rangle$ or $|V\rangle\xrightarrow{double\,QWP}|H\rangle$. Then, the photon reenters the storage loop from the $a_4$ port.
In this way, by controlling the on-off of the EOM, we can make the entering photon circulate in the storage loop to achieve the purpose of storage, or readout the photon from the storage loop to the $a_{1}$ mode. The period for the photon traveling one circle in the storage loop is controlled by the length of the delay line.
In detail, we provide the basic principle of the QM as
\begin{eqnarray}
   |H\rangle^{in}_{a_1}&&\xrightarrow{PBS}|H\rangle_{a_2}\xrightarrow{EOM(OFF)}|H\rangle_{a_3}\xrightarrow{PBS}|H\rangle_{a_4} \nonumber\\
    &&\xrightarrow{double\, QWP}|V\rangle_{a_4}\xrightarrow{PBS}|V\rangle_{a_2}\xrightarrow{EOM(ON)}|H\rangle_{a_3} \nonumber\\
   &&\xrightarrow{loop\ with\ EOM\ turn\ on}\cdots\xrightarrow{double\, QWP}|V\rangle_{a_4} \nonumber\\
  && \xrightarrow{PBS}|V\rangle_{a_2}\xrightarrow{EOM(OFF)}|V\rangle_{a_3}\xrightarrow{PBS}|V\rangle^{out}_{a_1},\nonumber
\end{eqnarray}
\begin{eqnarray}
|V\rangle^{in}_{a_1}&& \xrightarrow{PBS}|V\rangle_{a_3}\xrightarrow{EOM(OFF)}|V\rangle_{a_2}\xrightarrow{PBS}|V\rangle_{a_4}\nonumber\\
&&\xrightarrow{double\, QWP}|H\rangle_{a_4}\xrightarrow{PBS}|H\rangle_{a_3}\xrightarrow{EOM(ON)}|V\rangle_{a_2} \nonumber\\
&&\xrightarrow{loop\ with\ EOM\ turn\ on}\cdots\xrightarrow{double\,QWP}|H\rangle_{a_4}\nonumber\\
&&\xrightarrow{PBS}|H\rangle_{a_3}\xrightarrow{EOM(OFF)}|H\rangle_{a_2}\xrightarrow{PBS}|H\rangle^{out}_{a_1}.
\end{eqnarray}
It can be found that after the photon exiting the QM, its polarization will be flipped. In this way, after the photon is read out, we can add a half wave plate (HWP) to recover its polarization feature.

\begin{figure}[htbp]
    \begin{center}
    \centerline{ \includegraphics[width=9cm,angle=0]{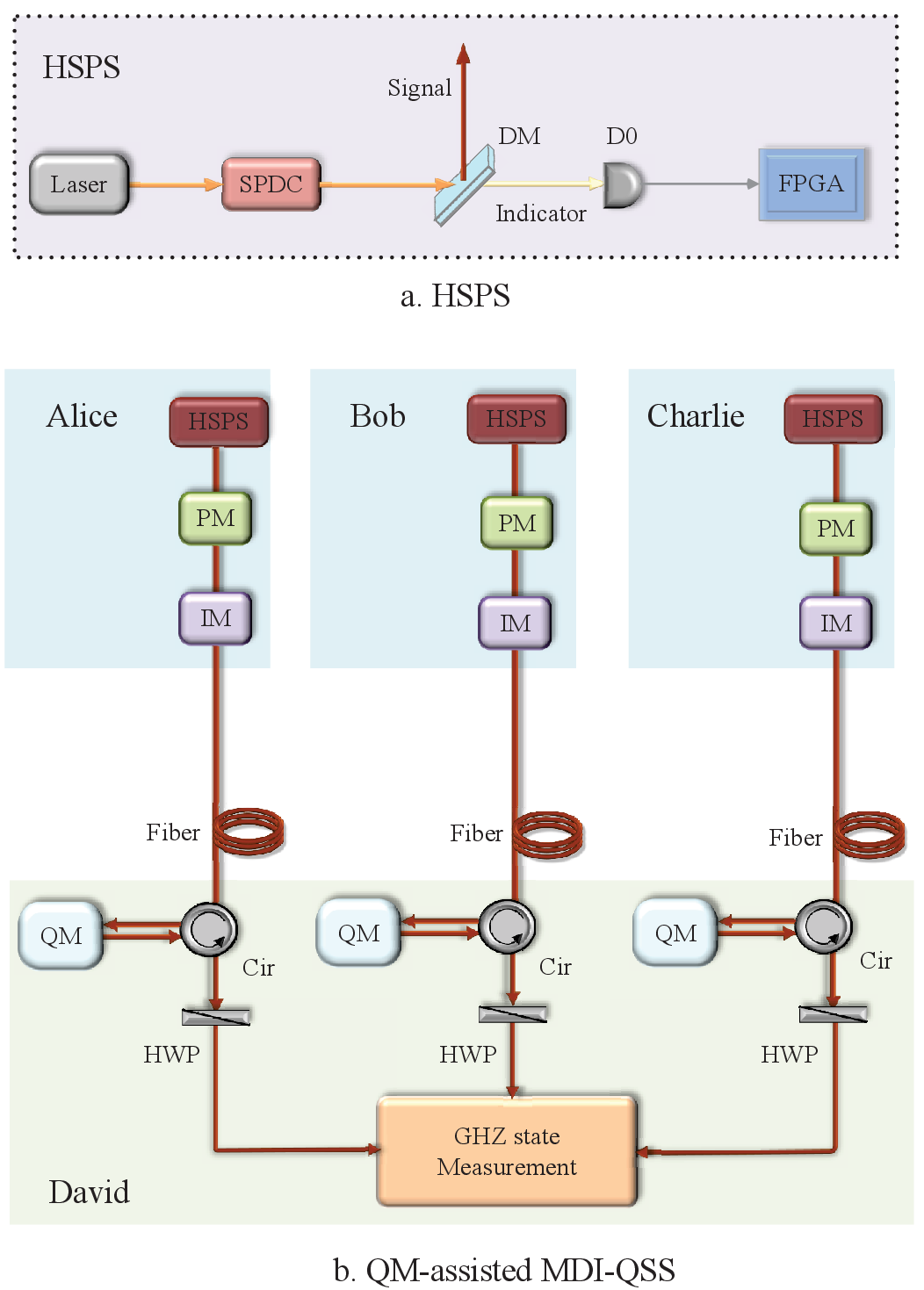}}
        \caption{(a) Structure diagram of the heralded-single-photon source (HSPS). Here, SPDC and DM represent spontaneous parametric down-conversion source and dichroic mirror, respectively. D0 is the single-photon detector. The event corresponding to D0's click is recorded by the signal control system FPGA. (b) The schematic diagram of the QM-assisted MDI-QSS protocol. Three users use the HSPSs to generate heralded single photons. The polarization modulator (PM) is used to encode the photons. The intensity modulator (IM) is used to modulate the intensity of the photon pulse. $QM_A$, $QM_B$ and $QM_C$ represent the QMs with the structure of Fig. 1(b), which store the photons from Alice, Bob and Charlie, respectively. The photon pulse enters and exits the QM through the circulator (Cir).}
        \label{fig:2}
    \end{center}
\end{figure}

Then, we introduce the basic principle of the HSPS. As shown in Fig. \ref{fig:2}(a), one passes a WCP laser to pump an SPDC crystal, splitting a single photon
to two photons probabilistically. One photon is in the signal path and the other photon is in the indicator path. The user detects the photon in the indicator path with the photon detector D0. The responses of D0 will herald the existence of the photon in the signal path. The event of D0's click is recorded by FPGA.

\subsection{The QM-assisted MDI-QSS protocol}
Here, we start to explain our QM-assisted MDI-QSS protocol, whose basic principle is shown in Fig. \ref{fig:2}(b). There are four participants in our protocol, i.e., the dealer Alice, the players Bob and Charlie, and the untrusted measurement party David, who is located in the middle node among the three users.

\emph{\textbf{Step 1 Single photon preparation.}}
Alice, Bob and Charlie each employ an HSPS to randomly generate the heralded single photon states. The detection result of $D0$ in each HSPS is processed by the FPGA.

\emph{\textbf{Step 2  Key coding.}}
After the heralded photon state is prepared, each of the three parties randomly encodes the single photon in Z basis or X basis by the polarization modulator (PM). The encoding rule can be described as follows. $|H\rangle$ and $|+\rangle$ represent the classical key bit 0, while $|V\rangle$ and $|-\rangle$ represent the classical key bit 1. After the encoding, each user uses the intensity modulator (IM) to randomly modulate photon pulse into the signal state or the decoy state.

\emph{\textbf{Step 3  Photon transmission.}}
 Alice, Bob and Charlie send the encoded photon pulses to the measurement party David.

\emph{\textbf{Step 4  Photon synchronization.}}
The measurement party consists of three QMs and a GHZ state measurement module. Since it is difficult for the photons sent by three users to reach the measurement module at the same time, a photon which arrives at David will be stored in the QM. By controlling the on-off of the EOM, the photon arriving first constantly circulates in the QM until the three photons from three users all arrive at the QMs. The optical storage loop is synchronized to the repetition rate of the pump laser. It is noted that the FPGA in each communication party controls the stored photon number of QM according to the detection result of $D0$. If the QM that has stored a photon receives a signal from the FPGA that a new photon is generated at the corresponding HSPS, the stored photon will be discarded and the new photon will be stored in the QM. When all the three photons from three users arrive at the QMs, the three photons will be read out from the QMs.

\emph{\textbf{Step 5  GHZ state measurement.}}
After the three photons leaving QMs, they each pass through an HWP to recover the initial states. Then, the photons are sent to a measurement module for the GHZ state measurement. The GHZ measurement module has the structure in the  Fig. \ref{fig:1}(a) \cite{GHZ}, which can only distinguish two GHZ states $|\Phi^{\pm}\rangle=\frac{1}{\sqrt{2}}(|HHH\rangle\pm|VVV\rangle)$. When the GHZ measurement fails, David announces the failure through a classical channel and the three users discard their encoded bits. When the GHZ state measurement is successful, David announces the measurement results and the three users preserve their encoded bits.

\emph{\textbf{Step 6 Security checking and raw key generation.}}
Alice, Bob and Charlie announce the generation basis for their preserved bits, respectively.
When all the three photons for the GHZ state measurement are generated in $Z$ basis, the photons are used for security checking. When the three photons for the GHZ state measurement are all generated in $X$ basis, the photons are used for generating the raw keys. From Tab. \ref{tab:1}, when all the three photons are generated in X basis, if the GHZ state measurement result is $|\Phi^+\rangle$, the users can obtain $k_A=k_B\oplus k_C$ and if the GHZ state measurement result is $|\Phi^-\rangle$, the users can obtain $k_A\oplus1=k_B\oplus k_C$. In this case, the encoded bits of Bob and Charlie are preserved as their raw key bits. If all the three photons from Alice, Bob and Charlie are generated in Z basis and the GHZ state measurement is successful, the users can obtain the correspondence of their encoded key bits ($k$) as $k_A=k_B=k_C$. In this case, Alice, Bob, and Charlie announce their encoded keys for the security checking. If their encoded keys do not meet $k_A=k_B=k_C$, they can deduce that an error occurs.
After all the measurements of security checking, the users estimate the total quantum bit error rate (QBER). If the value of QBER is higher than the tolerable threshold, the users ensure that the key generation process is not secure, so that they have to discard the generated raw key bits and recheck the quantum channels. If the value of QBER is lower than the tolerable threshold, the users ensure that the key generation process is secure. They continue to the next step.

\emph{\textbf{Step 7 Secure key generation.}}
Above steps are repeated until Bob and Charlie preserve enough raw key bits. Then, the users perform the error correction and private amplification on the obtained raw key bits, resulting in a series of secure key bits. Finally, Charlie announces his raw key bits and Bob can deduce Alice's key bit as the secure keys by combining Charlie's and his own key bits.

\section{Theoretical simulation}
\subsection{The successful probability of the three-photon synchronization}
Here, we show the theoretical simulation of the QM-assisted $M$-synchronized HSPSs \cite{yuanshi}. For simplicity, we suppose that the SPDC source emits the $n$-photon pulse with the probability of  $P_{\mu}(n)={\mu}^n/({1+\mu}^{n+1})$, which satisfies the heat distribution ($\mu$ is the average photon number) \cite{heat}.
 We define $P_d(k)$ as the probability that the trigger detector D0 of an HSPS clicks when an SPDC source generates a $k$-photon state, and $P_t(k'|k,j,j')$ as the probability that $k'$ of the $k$ photons emitted by the HSPS pass through the quantum channel, enter the QM at the $j'th$ time slot and exit at the $jth$ time slot. We also define $P_h(j)$ as the probability that an HSPS heralds at least one  photon in $j$ time slots. $P_d(k)$, $P_t(k'|k,j,j')$ and $P_h(j)$ can be calculated as
\begin{eqnarray}
P_d(k)&=&\sum_{l=1}^{k}\eta_D^l(1-\eta_D)^{k-l}\dbinom{k}{l},\nonumber\\
P_t(k'|k,j,j')&=&(T_cT_{QM}^{j-j'+1})^{k'}(1-T_cT_{QM}^{j-j'+1})^{k-k'}\dbinom{k}{k'},\nonumber\\
P_h(j)&=&1-[1-P_h(1)]^j,\nonumber\\
P_h(1)&=&\sum_{k=1}^{\infty}P_\mu(k)P_d(k).
\end{eqnarray}
  Here, $\eta_{D}$ is the detection efficiency of each detector, $T_c$ is the transmission efficiency of the quantum channel, and $T_{QM}$ denotes the storage efficiency of the QM for a delay time of $\tau$. It is noticed that $T_{QM}^{j-j'+1}$ corresponds to the photon enters the QM at the $j'th$ time slot and circulates in the QM for $j-j'+1$ times.

With the above definitions, the probability that $M$ photons emitted from $M$ HSPSs are projected synchronously into the measurement module is expressed as
\begin{eqnarray}
P_s(M)=P_l(1|1)^M+\sum_{j=2}^N\sum_{q=1}^M\dbinom{M}{q}P_l(1|j)^qP_e(1|j)^{M-q},\label{Ps}
\end{eqnarray}
where $P_l(k|j)$ is the probability that an HSPS initially heralds the generation of a $k$-photon state in the $j$th time slot ($N$ is the maximum number of storage time slots of each QM), and $P_e(k|j)$ is the probability of the HSPS heralding the generation of photons at least one time within $j-1$ time slots and then emitting a $k$-photon state at the $j$th time slot. In this way, $P_l(1|1)^M$ represents that each of the $M$ HSPSs generates a single photon in the first time slot. $P_l(1|j)$ and $P_e(1|j)$ can be written as
\begin{eqnarray}
P_l(1|j)&=&[1-P_h(j-1)]\sum_{k'=1}^{\infty}P_{\mu}(k')P_d(k')P_t(1|k',j,j),\nonumber\\
P_e(1|j)&=&\sum_{j'=1}^{j-1}\{[1-P_h(j-1-j')]\nonumber\\
&\times&\sum_{k'=1}^{\infty}P_{\mu}(k')P_d(k')P_t(1|k',j,j')[1-P_h(1)]\}\nonumber\\
         &+&P_h(j-1)\sum_{k'=1}^{\infty}P_{\mu}(k')P_d(k')P_t(1|k',j,j).
\end{eqnarray}

Here, we take the case of  $M=3$ for example. We use Eq. (\ref{Ps}) to calculate the successful probability $P_s(3)$ of the three-photon synchronous projection measurement as
\begin{eqnarray}
P_s(3)&=&P_l(1|1)^{3}+\sum_{j=2}^N [ 3P_l(1|j)P_e(1|j)P_e(1|j)\\
&+&3P_l(1|j)P_l(1|j)P_e(1|j)+P_l(1|j)P_l(1|j)P_l(1|j) ].\nonumber
\end{eqnarray}


\begin{figure}[t]
  \begin{center}
   \includegraphics[scale=0.65]{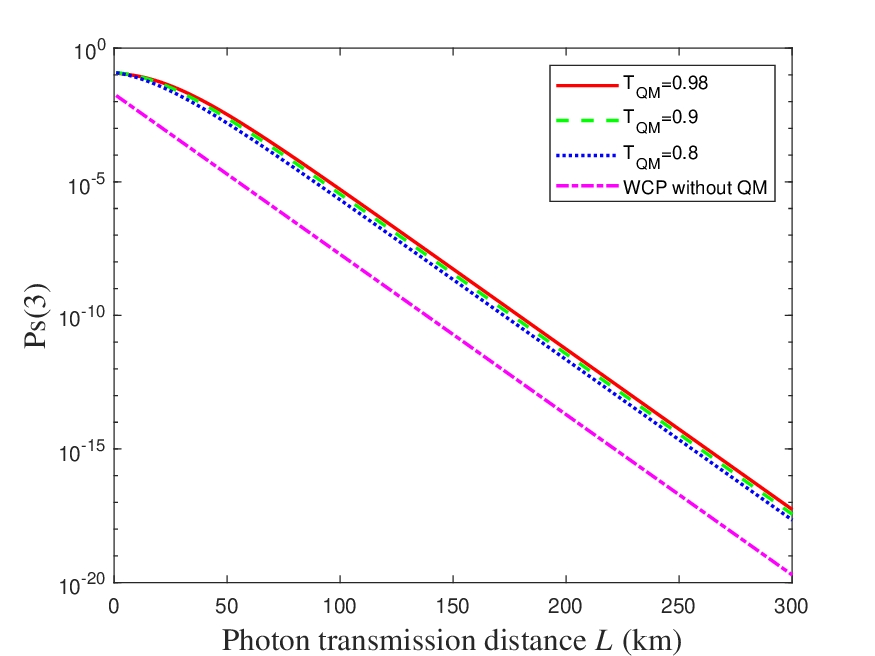}
   \end{center}
    \caption{The successful probability $P_s(3)$ of the three-photon synchronous projection measurement in our QM-assisted MDI-QSS protocol and the WCP-MDI-QSS protocol altered with the photon transmission distance ($L$). In our protocol, we fix $N=40$ and adjust $T_{QM}$=0.98, 0.9, 0.8, respectively. }
     \label{fig:3a}
\end{figure}
\begin{figure}[h]
    \begin{center}
     \includegraphics[scale=0.65]{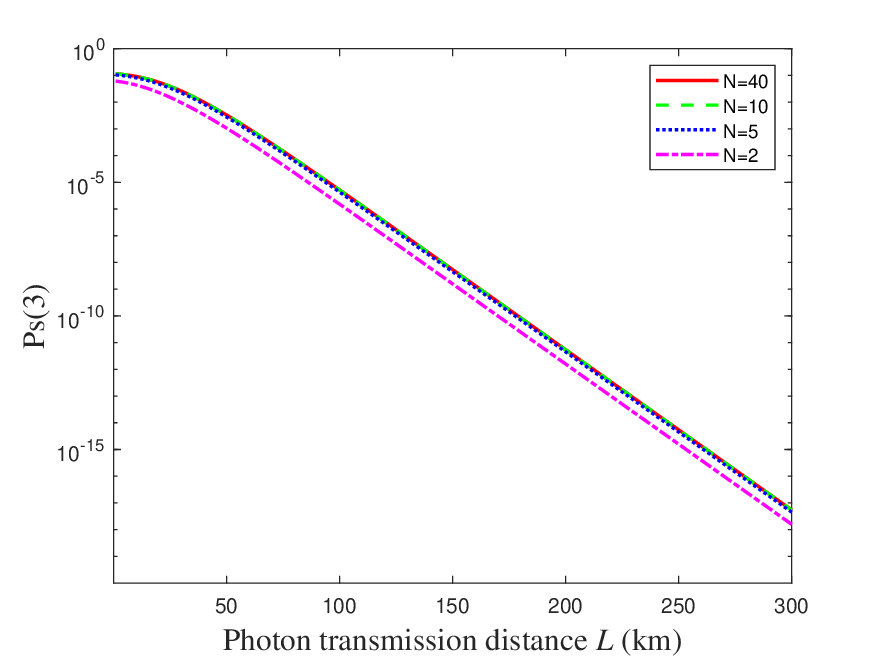}
     \end{center}
        \caption{The successful probability $P_s(3)$ of the three-photon synchronous projection measurement in our QM-assisted MDI-QSS protocol altered with the photon transmission distance ($L$). We fix $T_{QM}=0.98$ and adjust $N=$2, 5, 10, 40, respectively. }
        \label{fig:3b}
\end{figure}

We perform the numerical simulation of $P_s(3)$ altered with the photon transmission distance $L$ in Fig. \ref{fig:3a} and Fig. \ref{fig:3b}. The simulation parameters are shown in  Tab. \ref{tab:2}. In Fig. \ref{fig:3a}, we fix the maximum number of storage rounds $N=40$ and adjust the storage efficiency of the QM  $T_{QM}$=0.98, 0.9, 0.8, respectively. We also provide $P_s(3)$ of the previous WCP-MDI-QSS protocol without the QM \cite{MDI4}. From Fig. \ref{fig:3a}, it can be found that the adoption of QMs can effectively increase $P_s(3)$. In detail, when $T_{QM}=0.98$ and $L=200$ km, $P_s(3)$ of our QS-assisted MDI-QSS protocol is about $5.434\times10^{-12}$, which is about 280 times of that (about $1.928\times10^{-14}$) in the WCP-MDI-QSS protocol without the QM. Meanwhile, it is natural that $P_s(3)$ is influenced by the storage efficiency of each QM. $P_s(3)$ declines with the reduction of $T_{QM}$ from 0.98 to 0.8. The threshold of $T_{QM}$ is calculated as low as 0.183. If the $T_{QM}<0.183$,  $P_s(3)$ of our QM-assisted MDI-QSS protocol is lower than that of the WCP-MDI-QSS protocol \cite{MDI4}. In Fig. \ref{fig:3b}, we fix $T_{QM}$=0.98 and adjust $N=$2, 5, 10, 40, respectively. It can be found that under the condition of low value of $N$, i.e. $N<5$, the growth of $N$ would increase $P_s(3)$. However, under the condition of  $N\geq5$, the curves of $P_s(3)$ corresponding to $N=$5, 10, 40 almost overlap. It indicates that the further growth of $N$ would not increase $P_s(3)$ and the suitable value of $N$ is about 5 in practical applications.

Considering the imperfection of QM, we need to quantify the polarization error introduced by the QMs. According to Ref. \cite{wang}, when a polarization state $|V\rangle$ enters a QM, the state emitted from the QM can be defined as
\begin{eqnarray}
\rho_{out}&=&(1-e_{b})[(1-e_q)|V\rangle\langle V|+e_q|H\rangle\langle H|]\nonumber\\
&+&e_b(|H\rangle\langle H|+|V\rangle\langle V|)/2,
\end{eqnarray}
where $e_q$ is the error probability of the QM system, and $e_b$ is the probability that the QM is occupied by a noisy (unpolarized) photon and successfully reads it out. Considering the practical value
of $e_b$ is usually quite small, we can obtain the storage fidelity of a polarized photon as $F \approx 1-e_q$ \cite{wang}.

\subsection{Secure key rate }
In the asymptotic limit, i.e., without considering the finite-size effect, the secure key rate of our QM-assisted MDI-QSS protocol with the phase post-selection can be given by \cite{MDI4,ps}
\begin{eqnarray}
R&\geq&\frac{1}{K^2}Q^X_{111}[1-H(e^{BZ}_{111})]\nonumber\\
&-&H(E^X_{\mu_a\mu_b\mu_c})fQ^X_{\mu_a\mu_b\mu_c},\label{R}
\end{eqnarray}
where $K$ is the number of the phase regions, $\mu_a$ ($\mu_b$, $\mu_c$) represents the average photon number of Alice's (Bob's, Charlie's) photon pulse.

We first define the total gain of the right (error) measurement results as $Q_{\mu_a\mu_b\mu_c}^{RX}$ ($Q_{\mu_a\mu_b\mu_c}^{EX}$). It is noteworthy that when the GHZ state measurement result is $|\Phi^+\rangle$ ($|\Phi^-\rangle$), the right correlation among three users' key codes under the X basis is $k_A=k_B\oplus k_C$ ($k_A\oplus1=k_B\oplus k_C$). In this way, the correlation $k_A\oplus1=k_B\oplus k_C$ ($k_A=k_B\oplus k_C$) is the error correlation.
 Suppose that $e_d$ represents the overall misalignment-error probability of the GHZ state measurement module, combining the storage fidelity $F$ of the QM, the overall gain $Q_{\mu_a\mu_b\mu_c}^X$ and the total quantum bit error rate $E_{\mu_a\mu_b\mu_c}^X$ in the X basis can be given by
\begin{eqnarray}
Q_{\mu_a\mu_b\mu_c}^X&=&Q_{\mu_a\mu_b\mu_c}^{RX}
+Q_{\mu_a\mu_b\mu_c}^{EX},\\
E_{\mu_a\mu_b\mu_c}^X&=&\frac{[1-(1-e_d)F]Q_{\mu_a\mu_b\mu_c}^{RX}}{Q_{\mu_a\mu_b\mu_c}^X}\nonumber\\
&+&\frac{(1-e_d)FQ_{\mu_a\mu_b\mu_c}^{EX}}{Q_{\mu_a\mu_b\mu_c}^X}.
\end{eqnarray}
The derivation process of $Q_{\mu_a\mu_b\mu_c}^{RX}$ ($Q_{\mu_a\mu_b\mu_c}^{EX}$) is as follows \cite{MDI4}.

When Alice, Bob and Charlie all choose X basis, we have
\begin{eqnarray}
Q_{\mu_a\mu_b\mu_c}^{RX}&=&\frac{K}{\pi^2}\int_{0}^{\frac{\pi}{K}}\int_{0}^{\frac{\pi}{K}}
 \left[\right.F_{1H}F_{2H}F_{3H}(1-F_{1V})(1-F_{2V})\nonumber\\
 &&(1-F_{3V})+F_{1H}F_{2V}F_{3V}(1-F_{1V})(1-F_{2H})\nonumber\\
 &&(1-F_{3H})+F_{1V}F_{2H}F_{3V}(1-F_{1H})(1-F_{2V})\nonumber\\
&&(1-F_{3H})+F_{1V}F_{2V}F_{3H}(1-F_{1H})(1-F_{2H})\nonumber\\
&&(1-F_{3V})\left.\right]\,d\varphi\,d\phi,\nonumber\\
Q_{\mu_a\mu_b\mu_c}^{EX}&=&\frac{K}{\pi^2}\int_{0}^{\frac{\pi}{K}}\int_{0}^{\frac{\pi}{K}}\left[\right.
F_{1H}F_{2H}F_{3V}(1-F_{1V})(1-F_{2V})\nonumber\\
&&(1-F_{3H})+F_{1H}F_{2V}F_{3H}(1-F_{1V})(1-F_{2H})\nonumber\\
&&(1-F_{3V})+F_{1V}F_{2H}F_{3H}(1-F_{1H})(1-F_{2V})\nonumber\\
&&(1-F_{3V})+F_{1V}F_{2V}F_{3V}(1-F_{1H})(1-F_{2H})\nonumber\\
&&(1-F_{3H})\left.\right]\,d\varphi\,d\phi,
\end{eqnarray}
where  $F_{1H}$, $F_{1V}$, $F_{2H}$, $F_{2V}$, $F_{3H}$ and $F_{3V}$ are the click probabilities of the detector D1H, D1V, D2H, D2V, D3H and D3V, respectively. Here, the parameters $\phi=\theta_a-\theta_b$ and $\varphi=\theta_a-\theta_c$ ($\theta_a$, $\theta_b$ and $\theta_c$ represent the random phases of the photons prepared by Alice, Bob and Charlie, respectively). The click probabilities can be calculated as
\begin{eqnarray}
F_{1H}&=&1-(1-p_d)e^{-(\frac{\mu_a\eta_a+\mu_b\eta_b}{4}+\frac{\sqrt{\mu_a\eta_a\mu_b\eta_b}}{2}cos{\phi})},\nonumber\\
F_{1V}&=&1-(1-p_d)e^{-(\frac{\mu_a\eta_a+\mu_b\eta_b}{4}-\frac{\sqrt{\mu_a\eta_a\mu_b\eta_b}}{2}cos{\phi})},\nonumber\\
F_{2H}&=&1-(1-p_d)e^{-[\frac{\mu_b\eta_b+\mu_c\eta_c}{4}+\frac{\sqrt{\mu_b\eta_b\mu_c\eta_c}}{2}cos(\varphi-\phi)]},\nonumber\\
F_{2V}&=&1-(1-p_d)e^{-[\frac{\mu_b\eta_b+\mu_c\eta_c}{4}-\frac{\sqrt{\mu_b\eta_b\mu_c\eta_c}}{2}cos(\varphi-\phi)]},\nonumber\\
F_{3H}&=&1-(1-p_d)e^{-(\frac{\mu_a\eta_a+\mu_c\eta_c}{4}+\frac{\sqrt{\mu_a\eta_a\mu_c\eta_c}}{2}cos{\varphi})},\nonumber\\
F_{3V}&=&1-(1-p_d)e^{-(\frac{\mu_a\eta_a+\mu_c\eta_c}{4}-\frac{\sqrt{\mu_a\eta_a\mu_c\eta_c}}{2}cos{\varphi})},
\end{eqnarray}
where $\eta_a$, $\eta_b$ and $\eta_c$ are the overall detection
efficiencies of Alice, Bob and Charlie, respectively, and $p_d$ is the background count rate. In our protocol, we assume that the overall detection
efficiency of each user is equal. We can obtain $\eta_a=\eta_b=\eta_c=\eta_d\times\sqrt[3]{P_s(3)}$, where $\eta_d$ is the detection efficiency of the GHZ measurement module. $10^{-\alpha L/10}$ ($\alpha=0.2$ dB/km) is the channel transmission efficiency.

By referring to Ref. \cite{MDI4}, we can model the gains and the error rates with the two-intensity decoy-state (vacuum $+$ decoy state) MDI-QSS method. Assume that Alice, Bob and Charlie each have three
 identical intensities ($\mu$, $\omega$, $0$) in their state preparation, where $\mu_A=\mu_B=\mu_C=\mu$, $\omega_A=\omega_B=\omega_C=\omega$. We can estimate the lower bound of the yield
 for the single-photon pulses ($Y_{111}^{XL}$) and the upper bound of the bit error rate ($e_{111}^{BXU}$) as
\begin{eqnarray}
Y_{111}^{XL}&=&\frac{1}{P_{\mu}^2(1)P_{\omega}^2(1)
\left[P_{\mu}(2)P_{\omega}(1)-P_{\omega}(2)P_{\mu}(1)\right]}\nonumber\\\nonumber\\
&\times&\left[ P_{\mu}^2(1)P_{\mu}(2)\left(Q_{\omega\omega\omega}^X-P_{\omega}(0)Q_{\omega\omega o}^X-P_{\omega}(0)Q_{\omega o\omega}^X\right.\right.\nonumber\\
&-&P_{\omega}(0)Q_{o\omega\omega}^X+P_{\omega}^2(0)Q_{\omega oo}^X+P_{\omega}^2(0)Q_{o\omega o}^X \nonumber\\
&+&P_{\omega}^2(0)Q_{oo\omega }^X-P_{\omega}^3(0)Q_{ooo}^X\left.\right)-P_{\omega}^2(1)P_{\omega}(2)
 \left(Q_{\mu\mu\mu}^X
\right.\nonumber\\
&-&P_{\mu}(0)Q_{\mu\mu o}^X-P_{\mu}(0)Q_{\mu o\mu}^X-P_{\mu}(0)Q_{o\mu\mu}^X\nonumber\\
&+&P_{\mu}^2(0)Q_{\mu oo}^X+P_{\mu}^2(0)Q_{o\mu o}^X+P_{\mu}^2(0)Q_{oo\mu}^X\nonumber\\
&-&P_{\mu}^3(0)Q_{ooo}^X
 \left. \right) \left. \right],\\
 \nonumber\\
e_{111}^{BXU}&=&\frac{1}{P_{\omega}^3(1)Y_{111}^{XL}}
\left[\right.E_{\omega\omega\omega}^XQ_{\omega\omega\omega}^X-P_{\omega}(0)E_{\omega\omega o}^XQ_{\omega\omega o}^X\nonumber\\
&-&P_{\omega}(0)E_{\omega o\omega}^XQ_{\omega o\omega}^X
-P_{\omega}(0)E_{ o\omega\omega}^XQ_{o\omega\omega}^X\nonumber\\
&+&P_{\omega}^2(0)E_{\omega oo}^XQ_{\omega oo}^X+P_{\omega}^2(0)E_{o\omega o}^XQ_{o\omega o}^X\nonumber\\
&+&P_{\omega}^2(0)E_{oo\omega}^XQ_{oo\omega}^X-P_{\omega}^3(0)E_{ooo}^XQ_{ooo}^X
\left.\right].
\end{eqnarray}

Based on above formula derivation, we can obtain the lower bound of the gain $Q_{111}^{XL}=\frac{\mu}{1+\mu^{2}}Y_{111}^{XL}$ and the upper bound of the bit error rate $e_{111}^{BZU}=e_{111}^{BXU}$.
By taking the formula of $Q_{\mu_a\mu_b\mu_c}^X$, $E_{\mu_a\mu_b\mu_c}^X$, $Q_{111}^{XL}$, $e_{111}^{BZU}$ into Eq. (\ref{R}), we can obtain the lower bound of the secure key rate $R$.

\begin{table}[ht]
    \centering
    \vspace{-0.3cm}
    \setlength{\abovecaptionskip}{0.3cm}
    \setlength{\belowcaptionskip}{0.1cm}
    \setlength\tabcolsep{18pt} 
    \renewcommand\arraystretch{1.5}  
    \caption{ The list of the parameters used in the numerical simulation. Here, $p_d$ is the dark count rate of detectors; $e_q$ ($e_d$) denotes the misalignment probability of the QM (GHZ state measurement); $\eta_D$ ($\eta_d$) represents the detection efficiency of each detector in the HSPS (GHZ state measurement module); $f$ is the error correction inefficiency; $T_{QM}$ is the storage efficiency of the QM; $\alpha$ is the standard fiber loss coefficient.}
    \begin{tabular}{cccccc}
        \hline \hline
        \multicolumn{1}{c}{$p_d$} &\multicolumn{1}{c}{$e_q$}&\multicolumn{1}{c}{$e_d$}&\multicolumn{1}{c}{$\eta_D$} \\
       $10^{-7}$ & $1.5\%$ & $1.5\%$ & $93\%$ \\
       \hline
       $\eta_d$ &$f$ &$T_{QM}$ &$\alpha$ \\
       $93\%$ &$1.16$ &$98\%$ &0.2 dB/km \\
        \hline
        \hline
        \label{tab:2}
    \end{tabular}
\end{table}

 \begin{figure}[htbp]
    \begin{center}
        \includegraphics[width=9cm,angle=0]{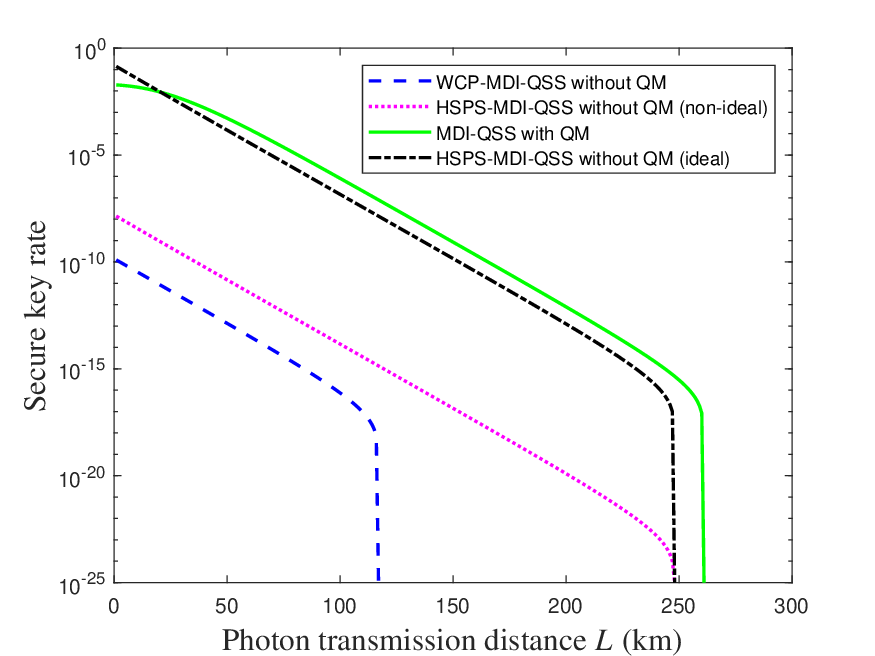}
        \caption{ The secure key rates of our QM-assisted MDI-QSS protocol and previous WCP-MDI-QSS and HSPS-MDI-QSS protocols without QM \cite{MDI4} versus the photon transmission distance ($L$). In our QM-assisted MDI-QSS protocol, we fix the total storage round and the storage efficiency of each QM as $N=40$ and $T_{QM}=$0.98, and the phase post-selection parameter as $K=8$. }
        \label{fig:4}
    \end{center}
\end{figure}

\begin{figure}[htbp]
    \begin{center}
         \includegraphics[width=9cm,angle=0]{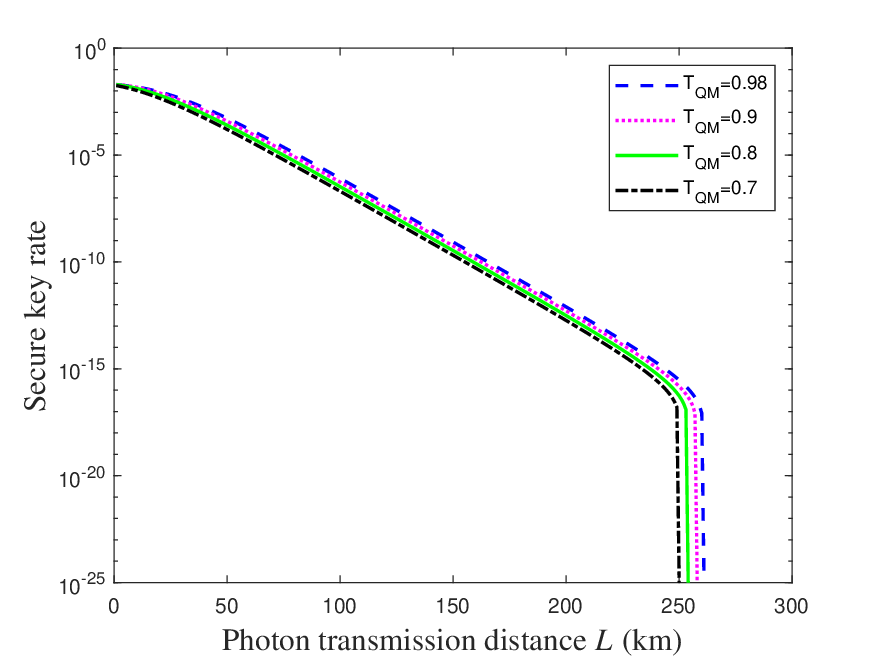}
        \caption{The secure key rate of our QM-assisted MDI-QSS protocol versus the photon transmission distance ($L$). Here, we fix $N$=40 and adjust $T_{QM}=$0.98, 0.9, 0.8, 0.7, respectively.  }
        \label{fig:5}
    \end{center}
\end{figure}

In Fig. \ref{fig:4}, we provide the secure key rates of our QM-assisted MDI-QSS protocol and the existing WCP-MDI-QSS and HSPS-MDI-QSS protocols without the QM \cite{MDI4} versus the photon transmission distance $L$, without considering the finite-size effect. The corresponding parameters are shown in Tab. II. In our QM-assisted MDI-QSS protocol, we set the total storage round is $N=40$, and the storage efficiency of QM as $T_{QM}=$0.98 based on previous studies \cite{yuanshi,QM}. The average photon numbers of the signal state and one decoy state is set as $\mu=0.005$ and $\omega=0.0005$, respectively, and the other decoy state is the vacuum state. As Ref. \cite{MDI4} only provides the key rate of the ideal HSPS-MDI-QSS protocol without considering the three-photon synchronization, we supplement the secure key rate of the non-ideal HSPS-MDI-QSS protocol by considering the probability of three-photon synchronization. It can be found that our QM-assisted MDI-QSS protocol has largely improved secure key rates compared with the WCP-MDI-QSS and HSPS-MDI-QSS (non-ideal) protocols. In detail, at the photon transmission distance of $L=100$ km, the secure key rate of our QM-assisted MDI-QSS is about 10 and 7 orders of magnitude higher than those of the WCP-MDI-QSS and HSPS-MDI-QSS (non-ideal) protocols, respectively. Meanwhile, our QM-assisted MDI-QSS protocol has longer maximal photon transmission distances. The maximal photon transmission distance of the WCP-MDI-QSS and HSPS-MDI-QSS (non-ideal) protocols are about 116 km and 248 km, respectively, while that of our QM-assisted MDI-QSS protocol achieves 261 km, about twice that of the WCP-MDI-QSS protocol. Here, we consider the symmetric model, say, the distances between each two communication parties are equal. In this way, the maximal communication distance between any two communication users of our protocol reaches about 452 km. The ideal HSPS-MDI-QSS protocol without the QM has about 7 orders of magnitude higher secure key rate than the non-ideal HSPS-MDI-QSS protocol but the same maximal photon transmission distance. When the photon transmission distance is relatively short ($L<21$ km), the secure key rate of our QM-assisted MDI-QSS is lower than that of the ideal HSPS-MDI-QSS protocol without QM. However, with the growth of the photon transmission distance ($L>21$ km), our QM-assisted MDI-QSS has higher secure key rate. At $L=100$ km, the secure key rate of our QM-assisted MDI-QSS protocol is about $8.129\times10^{-7}$, which is about 5.6 times of that in the ideal HSPS-MDI-QSS protocol (about $1.451\times10^{-7}$). Considering the HSPSs are excited
with a repetition rate of 10 GHz \cite{10G}, our QM-assisted MDI-QSS can achieve the secure key rate of 8129 bit/s.

We further investigate the effect of the QM storage efficiency on the secure key rate of our QM-assisted MDI-QSS protocol in Fig. \ref{fig:5}. It can be found that the maximal photon transmission distance can achieve 250 km, 254 km, 258 km, and 261 km corresponding to $T_{QM}=$0.7, 0.8, 0.9, 0.98, respectively. Meanwhile, the decline of  $T_{QM}$ would slightly reduce the secure key rate. By decreasing $T_{QM}=$ from 0.98 to 0.7, the secure key rate decreases from $8.129\times10^{-7}$ to $2.025\times10^{-7}$.

\section{Discussion and conclusion}

 In our work, we propose a QM-assisted MDI-QSS protocol, which employs three QMs to synchronize three HSPSs to efficiently generate three simultaneous single photons. Here, we discuss the experimental realization of our QM-assisted MDI-QSS protocol. The QM constructed with all-optical, polarization-insensitive storage loop is the key element of our protocol.  The all-optical storage loops were employed in Ref. \cite{QM} for the experimental generation of four-photon and six-photon GHZ states with the help of the entanglement swapping. It is interesting to compare the all-optical storage loop QM
with the atomic QMs for the polarization qubit, such as those based on electrically
induced transparency (EIT) in cold caesium \cite{QM3} and rubidium \cite{QM4} ensembles, or atomic frequency combs (AFC)
in neodymium \cite{QM1} and europium \cite{QM2}. Ref. \cite{QM} experimentally achieved the storage efficiency of the all-optical storage loop QM as $T_{QM}=91\%$ and a lifetime of 131 ns, corresponding to around 11 round-trips. Especially, in the region up from 11 to 20 round-trips, i.e., 20
multiplexed sources, the average storage fidelity of the loop QM only decreases
from 99.7\% to 98.5\%. Those measurement were
taken for photons with a central wavelength of 1550 nm and a
bandwidth of 0.52 THz. In this way, the loop QM  has better performance than the atomic QMs in terms of
bandwidth, storage efficiency, and noise resistance. Although the storage loop QM has lower memory lifetime than the atomic QMs \cite{QM3,QM4} (loop QM: 131 ns, the atomic QMs in Refs.\cite{QM3,QM4}: $\sim$ $\mu$s), the lifetime of 131 ns is sufficient for the Bell state measurement and GHZ state measurement. Moreover, the loop QM can operate at
any given wavelength in principle with only the minor adaptions. In contrast, the atomic QMs govern the operation wavelength
of the photon based on the atomic level structure of the
underlying material system. Benefiting from the promising all-optical, polarization-insensitive storage loop QM, our protocol can realize the feasible and high-efficient MDI-QSS.

 With the QM-assisted synchronization operations, our QM-assisted MDI-QSS protocol can increase the maximal photon transmission distance. Interestingly, there are some other possible approaches to further increase the maximal photon transmission distance, such as the adoption of the quantum repeater (QR) \cite{repeater0,repeater1,repeater2,repeater3,repeater4}, especially the third-generation of QR. The third-generation of QR counteracts the errors resulted from both photon loss and imperfect operations by employing quantum error correction (QEC) codes, in which the damaged QEC codes can be recovered to the complete ones as long as the error rate is lower than the fault tolerance \cite{repeater1,repeater4}. Combining our QM-assisted MDI-QSS protocol with QR may be a promising way to further increase its secure communication distance and realize the long-distance MDI-QSS. This approach will be investigated in our future works.

In conclusion, we propose a high-efficient QM-assisted MDI-QSS protocol, which employs the QM-assisted synchronization system of three HSPSs for efficiently generating three simultaneous single-photon states. The QM constructed with all-optical, polarization-insensitive storage loop has superior performance in terms of bandwidth, storage efficiency, and noise performance. Moreover, it is feasible under current experiment conditions. The adoption of the QM-assisted synchronization system can largely increase the successful probability of the three-photon synchronous projection measurement and thus increase the secure key rate and photon transmission distance of the MDI-QSS protocol. We perform the numerical simulation of the secure key rate in the symmetric model without considering the finite-size effect. From the simulation results, our MDI-QSS protocol has the maximal photon transmission distance of 261 km, which is longer than those of the WCP-MDI-QSS protocol without QM (116 km) and HSPS-MDI-QSS protocol without QM (248 km), respectively. In this way, the maximal distance between each two communication users of our QM-assisted MDI-QSS protocol reaches about 452 km.  When the photon transmission distance is 100 km, the secure key rate of our QM-assisted MDI-QSS is $8.129\times10^{-7}$, which is about 10 and 7 orders of magnitude higher than those of the WCP-MDI-QSS and HSPS-MDI-QSS (non-ideal) protocols without the QM, respectively. Our protocol provides a promising way for implementing the high-efficient MDI-QSS in the near future.

\section*{Acknowledgement}
This work was supported by the National Natural Science Foundation of
China (Grant Nos. 12175106 and 92365110), and the Postgraduate Research \& Practice Innovation Program of Jiangsu Province under Grant No. KYCX23-0989.

\end{document}